\documentclass{research4cacm}

\usepackage{enumitem}
\usepackage{csquotes}
\usepackage{xcolor,soul}
\usepackage{setspace}
\usepackage{array}
\usepackage{url}
\MakeOuterQuote{"}

\CopyrightYear{2018}

\definecolor{dark-red}{rgb}{0.6,0.15,0.15}
\definecolor{dark-green}{rgb}{0.15,0.4,0.15}
\definecolor{medium-blue}{rgb}{0,0,0.5}
\usepackage[plainpages=false,   
            pdfpagelabels,      
            unicode,           	
            colorlinks,    		
            linkcolor={dark-red},
            citecolor={dark-green},
            urlcolor={medium-blue}
            ]{hyperref}

\definecolor{lightyellow}{RGB}{255, 243, 178}
\definecolor{lightred}{RGB}{250, 157, 157}
\definecolor{lightblue}{RGB}{125, 211, 240}

\newcommand{\notcite}[1]{}

\begin{document}

\title{A large-scale comparative study\\of beta testers and standard users
\thanks{This is the author's final version. The final publication is available at ACM via \href{https://doi.org/10.1145/3173570}{doi.org/10.1145/3173570}. Further paper details available at \href{https://crocs.fi.muni.cz/papers/cacm2018}{crocs.fi.muni.cz/papers/cacm2018}.}
}
\subtitle{Case study of a software security system}

\numberofauthors{6}
\author{
\alignauthor
Vlasta Stavova\\
    \affaddr{Faculty of Informatics}\\
    \affaddr{Masaryk University}\\
    \affaddr{Czech Republic}
    \email{\hspace{-0.7cm}vlasta.stavova@mail.muni.cz}
\alignauthor
    Lenka Dedkova\\
    \affaddr{Faculty of Social Studies}\\
    \affaddr{Masaryk University}\\
    \affaddr{Czech Republic}\\
    \email{ldedkova@fss.muni.cz}
\and
\alignauthor
    Martin Ukrop\\
    \affaddr{Faculty of Informatics}\\
    \affaddr{Masaryk University}\\
    \affaddr{Czech Republic}\\
    \email{mukrop@mail.muni.cz}
\alignauthor
    Vashek Matyas\\
    \affaddr{Faculty of Informatics}\\
    \affaddr{Masaryk University}\\
    \affaddr{Czech Republic}\\
    \email{matyas@fi.muni.cz}
}

\maketitle

Beta testing is an important phase of product development, where a sample of target users (potential adopters) try a product before its official release. It is practically ubiquitous: everywhere from medicine to software development, participants test and troubleshoot products to help improve their functioning and avoid defects.

Should you care who the beta testers actually are? We believe you definitely should. In order to generalize beta testing outcomes, the population of testers must be as representative of the final customers as possible. If this is not the case, results of beta testing can be heavily biased and can fail to capture important product flaws. In other words, different purposes of beta testing demand different beta testers.
We feel that these aspects are strongly underestimated in software beta testing -- companies often use any beta testers available, without a proper selection, and afterwards without analyzing to what extent were the beta testers comparable to the population of (targeted) end users.
Earlier on, it was easier to know product beta testers well~\cite{dolan1993maximizing}. This is not true in the software industry anymore, thanks to the Internet and the pace of releasing updates and new versions.
We strongly believe that firms should pay attention to their beta testers and select them wisely, because appropriate beta testing is more efficient and more economical.

Costs of poor beta testing were apparent from the very beginning of software beta testing. An early example from the 1990s is a software company that chose only one site for beta testing \cite{dolan1993maximizing}. Based on test results, developers made several changes. Since the beta testers represented only a specific sub-population of intended users, the product became so customized that it could not be marketed to other institutions. 
In 2012, the Goko company released a web portal for developing multiplayer games. They used too small pool of beta testers, so they did not find a serious bug connected with the site load \cite{betatesting}. And beta testing is not only about bug-hunting: there are also product support enhancement and marketing benefits~\cite{dolan1993maximizing}.

We present the case study results of a comparison between beta testers and standard users of an online security software. Altogether, we analyzed records of over 600\,000 participants who gave consent with being part of this study. We focused on their similarities and differences to discover whether beta testers represented the standard users well enough. Despite the fact that alpha testers are well-described in literature, as far as we know, no larger field study focused on a comparison of beta testers and standard users was published. Thus, we present what we believe to be the first public large-scale comparison of beta testers and standard users.

As a result of our comparison, we elaborated whether companies should be more selective about beta testers or just go with the intuitive approach `the more testers the better the final results'.
We do not aim to investigate goals or parameters and conditions of beta testing.

We outlined three main research issues to be examined for our subsamples of beta testers and standard users:
\begin{enumerate}
  \item Technology: \textit{Do the subsamples have similar profiles with respect to the hardware and operating system?}
\item Demography: \textit{Do the subsamples have similar age, gender and education profiles? What about cultural background distribution?}
\item Computer self-efficacy and privacy perception: \textit{Do the subsamples see themselves as equally skilled regarding their use of computers? Do they perceive their data as similarly safe?}
\end{enumerate}

The following two sections review previous works in the field and describe methods and the analyzed data. Following three sections then map the results for the three research issues stated above. The last two sections discuss study limitations and conclude with contributions and actionable takeaways.

\section{Study rationale}

Testing represents 30-50\% of software development costs~\cite{biffl2006value} and about 50\% of development time~\cite{pan1999software}. There are many testing phases, the first round(s) usually utilizing alpha testers. Their number is limited by a company size, and even for big companies it is impossible to duplicate the myriad of possible hardware/software configurations. Whereas alpha testers are typically company employees, beta testers are the first product users outside the company. Their feedback can greatly influence the product before it is used by standard customers.

Thanks to the Internet, thousands of beta testers with different devices and practices can report their feedback on a product before its full launch. An additional benefit of beta testing lies in involving international aspects. Since beta test participants can come from different locations, localization issues (such as language, currency, culture or local standards) can be detected and reported~\cite{betatesting}. Furthermore, cultural background also affects the perceived usability~\cite{wallace2009effect}. Therefore, beta testers bring huge benefits by detecting potential hardware conflicts and performing usability checking.

Our work summarizes a large-scale case study on beta testers and standard users of an online security software. While many alpha or beta testing studies have been published, the idea of comparing beta testers and standard users has been rarely tackled before. Mantyla et al. investigated the related question \textit{Who tested my software?}~\cite{mantyla2012tested}, but their study was limited to three companies' employees. Other studies~\cite{kanij2015empirical, merkel2010does} provide some insights into the software tester population, yet are mainly based on specific sub-populations, such as people interested in testing, users of specialized forums and LinkedIn, or companies' employees, so a selection bias may occur.

We compare beta testers and standard users in a number of aspects.
Firstly, we focus on technology. Having similar devices with regard to the technical aspects (hardware, OS, etc.) is the basic requirement for successful software beta testing. Since physical environment is one aspect that influences usability testing~\cite{sauer2010influence}, the device used to test an application may influence its usability too. For example, security software running in background can decrease the perceived overall performance of the machine and thus the perceived usability. Therefore, participants with a low-end hardware may encounter different usability issues than those with a high-end one. Usually, beta testers are thought of as being the problem solvers, or early adopters~\cite{betatesterstype} with most recent and hi-tech hardware. Therefore, many issues could stay unnoticed during beta testing.

Secondly, we examine user demographics. Existing research shows that users' IT-related behavior is largely affected by their gender, age, education and cultural background. For example, a greater rate of computer use and online activities was associated with lower age, higher education and being male~\cite{2015world}. The differences in IT usage are also related to the country of origin~\cite{ono2007digital}. The countries differ in the development of information society, leading to varying access opportunities and creating digital disparities between nations~\cite{chinn2010ict,cuervo2006multivariate}. As a result, there could be nations with more computer savvy populations and/or populations more inclined to use free software (even though still in beta). For example, anecdotal evidence has it that the Japanese move towards emerging technologies more slowly than other nations~\cite{2011japan}.

Thirdly, the different patterns of Internet/computer usage are associated with other individual characteristics, such as users' computer self-efficacy and privacy perceptions. Computer self-efficacy~\cite{compeau1995computer} reflects the extent to which the user believes he/she is capable of working efficiently with a computer. Users with a higher computer self-efficacy tend to use the computer more~\cite{compeau1995computer}, adopt new technology faster~\cite{hill1986communicating,venkatesh2003user} and have better performance in computer-related tasks~\cite{downey2009accurately}. Regarding privacy perception, marketing research constantly shows how consumer online behavior (such as willingness to provide personal information or intention to use online services) are affected by their privacy concerns~\cite{malhotra2004internet,sheehan2002toward}. Since beta testing usually includes sharing one's system, localization or even personal information with a company, it may discourage users with higher privacy concerns or those who store more private data on their computers. However, this may be an important segment of end users population with distinct expectations from the final product.

\section{Methodology}

The study was conducted in cooperation with ESET, an online security software company with over 100 million users in more than 200 countries and territories\footnote{https://www.eset.com/int/about/}. Two samples were used for analyses: beta testers and standard users of ESET security software solution for Windows. 

ESET Beta program allows anybody to download the product beta version and become a public beta tester. Despite the fact that users fill out a questionnaire before they can beta test the product, the company does not use any criteria to select beta testers. The main role of testers is to report bugs and/or suggest improvements. ESET's beta testers are motivated by the opportunity to use a beta product for free and possibility to use the product sooner than standard users.

The sample of beta testers ($N$ = 87\,896) was collected from June to December 2015, the sample of standard users ($N$ = 536\,275) from January to March 2016.

Firstly, we collected anonymized system parameters for each installation including processor configuration, RAM size, operating system, country, and time spent on each installation screen. Countries were identified by GeoIP2\footnote{https://www.maxmind.com/en/geoip2-databases}. One data record represents one installation of the software.

Secondly, a questionnaire was introduced to users at the end of the installation process. Filling out a questionnaire was voluntary and we used no incentives other than stating that completing it will help ESET improve their products. Of the beta testers sample, 6\,008 users filled out at least one questionnaire item (7.800\%) and the same applied to 27\,751 standard users (5.560\%). Since we collected the data from installations in English, the questionnaire was also presented in English. The questionnaire was a source for demographic data and privacy perceptions. No identification data were collected.

\subsection{Data cleaning}

During the data cleaning, we first removed installations coming from ESET's internal IP space (0.282\% of the sample) to exclude company alpha testers. Further, since each data entry reflected one installation, there might have been duplicated entries from the same device. To prevent possible biases, we identified cases with the same combination of hardware specification and IP address, randomly selected one and deleted the rest (this removed 7.429\% of the data). 

As noted, the questionnaire was voluntary and only a sub-sample of users completed it. The whole questionnaire was presented on four screens. We used the time spent on each screen to clean the data -- those who spent less than 6 seconds on a screen with two items and those who spent less than 7 seconds on a screen with three items were considered as invalid and their data from respective screen were omitted from analyses ($N$ = 10\,151; 30.1\% of the questionnaire respondents).

The final cleaned sample for our study thus consists of 576\,170 installations on unique devices, including 29\,598 questionnaires with at least one answered item, see Table~\ref{tab:participants-numbers}.

\begin{table}[tbp]
\centering
\begin{tabular}{l||r|>{\raggedleft}p{1.3cm}|r}
& Unique devices & \multicolumn{2}{c}{Filled questionnaires} \\ \hline \hline
Beta testers & 77\,028 & 5\,514 & 7.158\% \\ \hline
Standard users & 499\,142 & 24\,084 & 4.825\% \\ \hline \hline
Total & 576\,170 & 29\,598 & 5.137\%
\end{tabular}
\caption{An overview of participant numbers (after data cleaning).}
\label{tab:participants-numbers}
\end{table}

\subsection{Analytical strategy}
To assess the differences between standard users and beta testers, we used the $\chi^2$ test (categorical data) and t-tests (interval data). Analyses on large samples typically show statistically significant results even for very small effects. When considering such results, it is important to interpret effect sizes rather than to rely on significance alone. Therefore, we calculated Cramer's~$V$ ($\varphi_c$) for categorical data and Cohen's~$d$ for interval data. For $\varphi_c$, the value of 0.1 is considered as small, 0.3 as medium and 0.5 as a large effect size, and for $d$, the respective values are 0.2, 0.5, and 0.8~\cite{cohen1988,field2002design}.

The fact that the questionnaire data came only from a sub-sample of users may imply a bias in the results (see the limitations). To gain an insight in the differences between the samples with and without the questionnaire, we compared these users with regard to the parameters available for all of them (platform information, CPU performance, RAM and OS version). The effect sizes of found differences were negligible ($\varphi_c$ < 0.034). Therefore we are confident that the questionnaire data are informative, despite being obtained from only a small subsample of users. 

\section{Technology}

First we look at the technological aspects. We inspect the hardware platform information (32/64-bit), CPU model, RAM size and OS version. 

\subsection{Hardware}

The platforms differ only slightly between the subsamples -- 35.3\% of beta testers use 32-bit systems while about 34.5\% of the standard users do so ($\chi^2$(1)= 20.998, $\varphi_c$ = -0.006, $p$ < 0.001, $N$ = 576\,170).

The CPU performance has been categorized into four groups (low-end, mid-low, mid-high, high-end) based on the PassMark \textit{CPU Mark} criterion~\cite{passmark}. The CPU name was matched against the PassMark online database. Since CPU names are not standardized, we were unable to assign the score in 3.040\% of the cases ($N_{noCpuMark}$ = 17\,514, proportionally distributed between beta testers and standard users).

\begin{figure}
\centering
\includegraphics[width=\columnwidth]{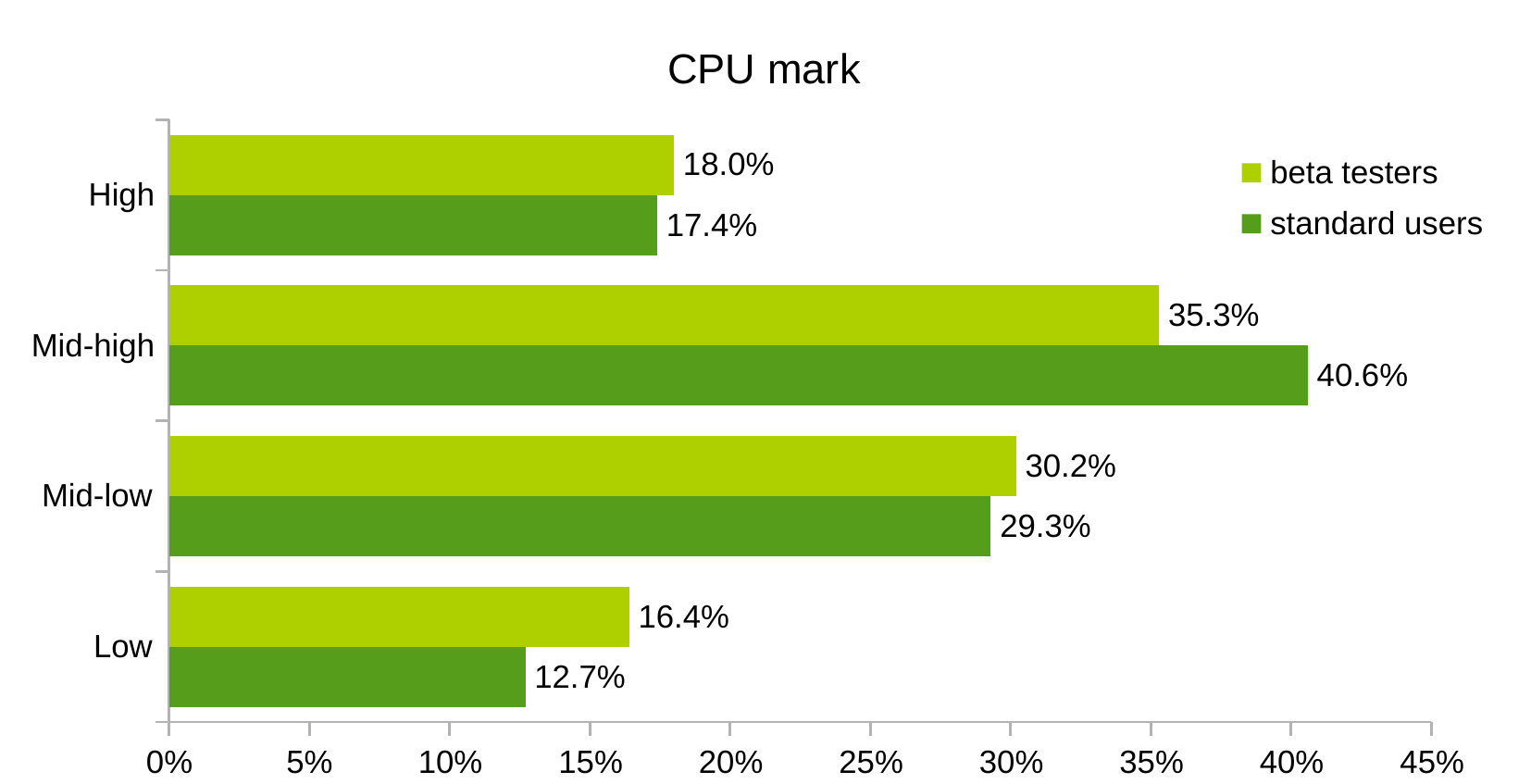}
\includegraphics[width=\columnwidth]{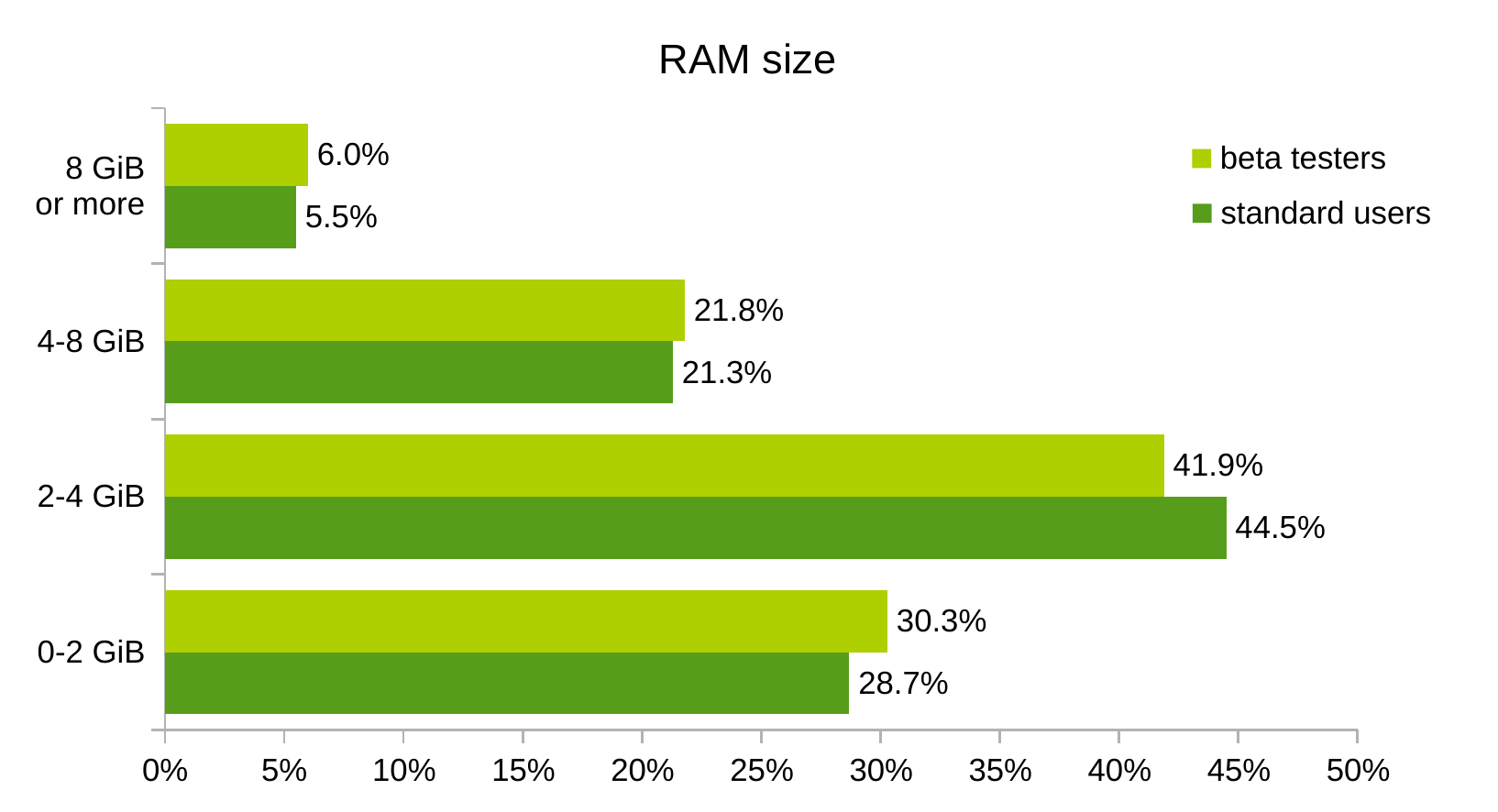}
\caption{Basic hardware characteristics for beta testers and standard users (read more in Study Limitations, same for the following figures).}
\label{fig:cpuram}
\end{figure}

Beta testers are more represented in the low-performance category and standard users in the mid-high category. The proportions are quite similar in mid-low and high-end categories (see Figure~\ref{fig:cpuram}). Although statistically significant, the effect size is small ($\chi^2$(3)= 1187.546, $\varphi_c$ = 0.045, $p$ < 0.001, $N$ = 576\,170). 

RAM size was grouped into 4 categories (0-2\,GB, 2-4\,GB, 4-8\,GB, 8\,GB and higher). Standard users' proportion is higher in the `2-4\,GB' category, while beta testers dominate in the lowest `0-2\,GB' category. The proportions in two highest categories are similar, see Figure~\ref{fig:cpuram}. The small effect size suggests the differences are negligible, despite being significant ($\chi^2$(3)= 206.926, $\varphi_c$ = 0.019, $p$ < 0.001, $N$ = 576\,170).

\subsection{Operating system}

\begin{figure}
\centering
\includegraphics[width=\columnwidth]{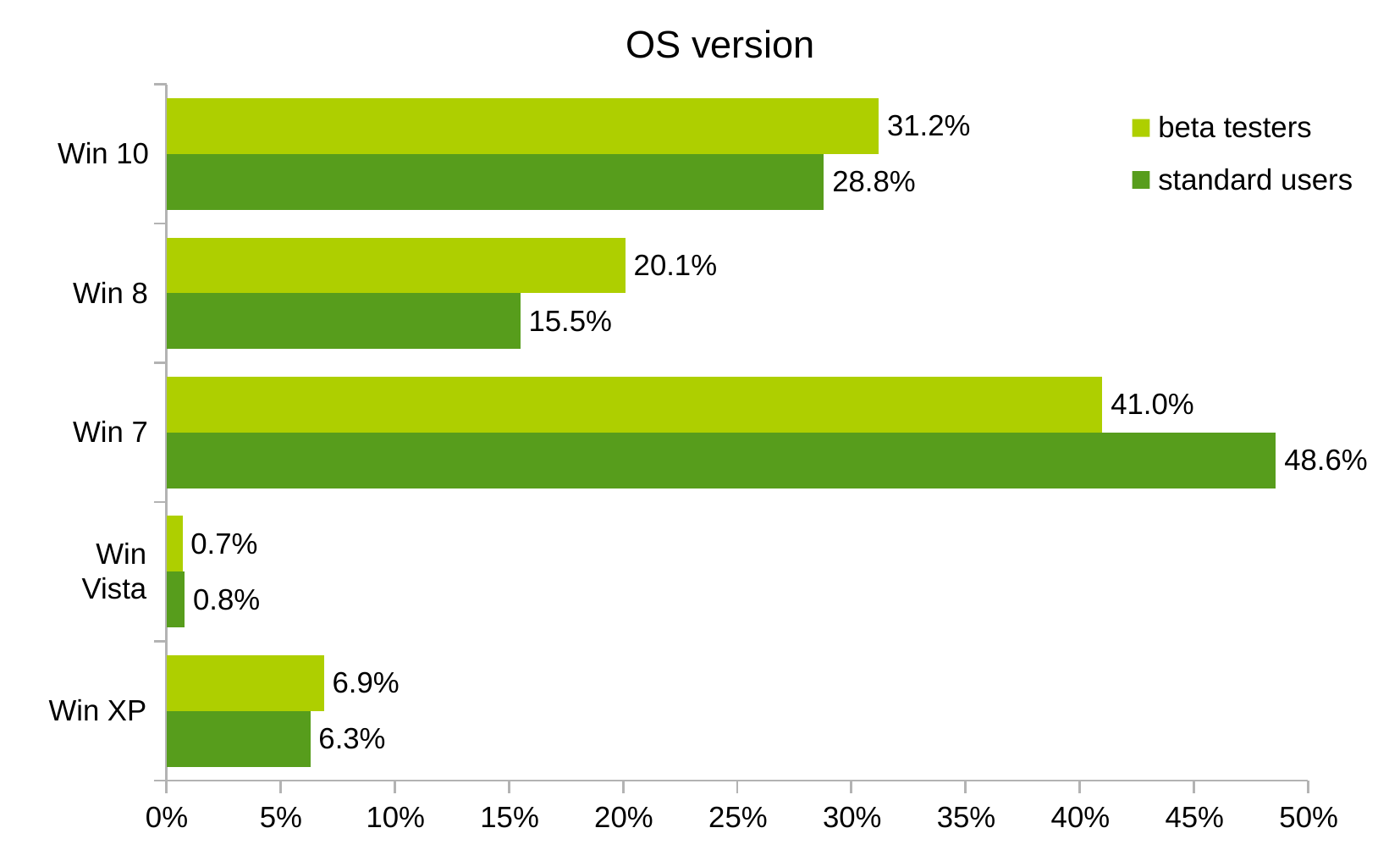}
\caption{Comparison of operating system version between beta testers and standard users.}
\label{fig:winversions}
\end{figure}

Beta testers prevail in the two newest OS versions (Windows 8, Windows 10), while standard users dominate in Windows 7, and have nearly equal representation in Windows Vista and XP, see Figure~\ref{fig:winversions}. The effect size is again small ($\chi^2$(2)= 1\,925.745, $\varphi_c$ = 0.058, $p$ < 0.001, $N$ = 575\,979). Other Windows versions (Windows 98, Windows 2000, etc.) were also marginally present but were omitted due to the extremely low counts (<0.001\%, $N_{otherWinVersions}$ = 191). Note that the study targeted only users of Microsoft Windows software.

Windows 10 was more often used by beta testers, even though their data were collected sooner -- i.e., standard users had more time to upgrade. This observation indicates that beta testers are often recruited from early adopters~\cite{betatesterstype}.

\pagebreak
\subsection{Specific configurations}
We were also interested in specific configurations of users' devices. We combined all four technological aspects together (OS platform, CPU performance, RAM size, OS). This led to the identification of 116 unique hardware+software combinations in the whole dataset with frequencies from 1 to 43\,519 (7.556\%). The sample of standard users included 114 combinations (i.e., there were two specific combinations found among beta testers' devices, which were not found among standard users), and the sample of beta testers included 102 combinations. However, the combinations not present in beta testers were also rather marginally present among standard users ($N_{onlyStandard}$ = 52, 0.010\%). We conclude that for almost every standard user in the sample, there was a beta tester in the sample with the same combination of examined parameters.

\section{Demography}
This section discusses participants' cultural and demographic profiles. We focus on the country of origin, gender, age and achieved education.

\subsection{Country of origin}

\begin{figure}
\centering
\includegraphics[width=\columnwidth]{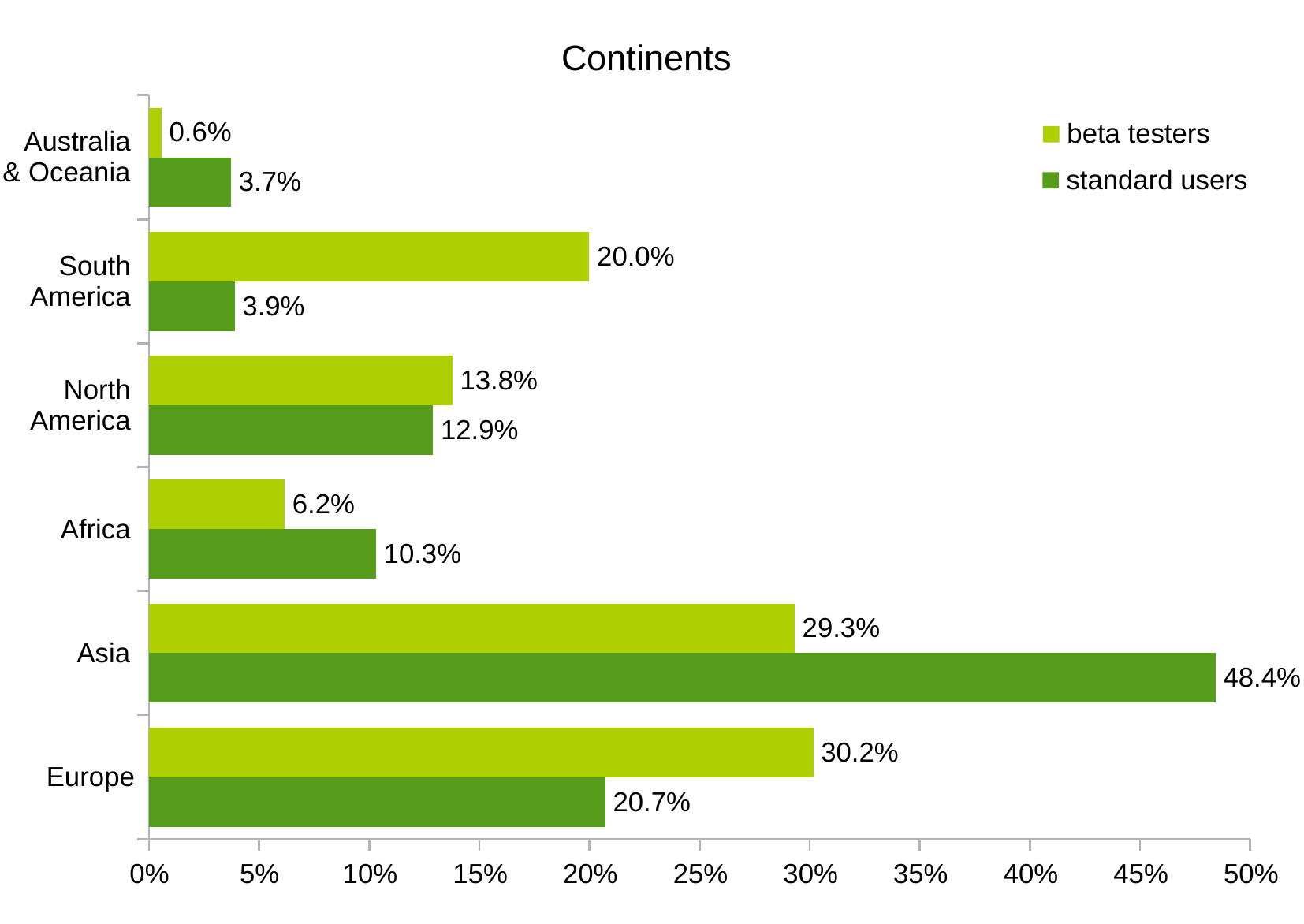}
\caption{The comparison of beta testers and standard users with respect to the continent they come from.}
\label{fig:continents}
\end{figure}

As noted in the Methodology section, the country was based on GeoIP2. This procedure failed to assign a country to 0.4\% cases ($N_{noCountry}$ = 2\,408). For a better overview, countries were grouped by continents and both subsamples were compared (see Figure~\ref{fig:continents}). We observed significant differences -- beta testers substantially dominate in South America and Europe, while standard users are more often based in Asia, Africa and Australia/Oceania ($\chi^2$(5)= 39\,049.72, $\varphi_c$ = 0.261, $p$ < 0.001, $N$ = 573\,538).

A detailed information regarding most represented countries can be seen in Table~\ref{tab:countries}. Only Iran, India, Egypt and the USA are among these most represented ones in both subsamples. 

These issues are now investigated in the company w.r.t. product localization and usability, where country differences are quite likely to play a role.
\begin{table}
\centering
\begin{tabular}{|p{1.8cm}rr|p{1.8cm}rr|} \hline
\multicolumn{3}{|c|}{\bfseries Standard users} & \multicolumn{3}{c|}{\bfseries Beta testers} \\
\bfseries Country & \bfseries N & \bfseries \% & \bfseries Country & \bfseries N & \bfseries \% \\ \hline
Iran & 81\,035 & 16.2 & Mexico & 5\,662 & 7.4 \\ \hline
USA & 50\,220 & 10.1 & Indonesia & 5\,117 & 6.6 \\ \hline
India & 26\,532 & 5.3 & Brazil & 4\,251 & 5.5 \\ \hline
Indonesia & 25\,959 & 5.2 & China & 4\,132 & 5.4 \\ \hline
UK & 25\,173 & 5.0 & Peru & 3\,422 & 4.4 \\ \hline
Egypt & 21\,649 & 4.3 & Russia & 3\,348 & 4.3 \\ \hline
Romania & 16\,582 & 3.3 & Ukraine & 2\,979 & 3.9 \\ \hline
Pakistan & 15\,831 & 3.2 & Spain & 2\,513 & 3.3 \\ \hline
Peru & 15\,280 & 3.1 & Egypt & 2\,393 & 3.1 \\ \hline
Philippines & 14\,904 & 3.0 & <unknown> & 2\,314 & 3.0 \\ \hline
South Africa & 13\,951 & 2.8 & Iran & 2\,306 & 3.0 \\ \hline
UAE & 11\,584 & 2.3 & India & 1\,771 & 2.3 \\ \hline
Thailand & 10\,719 & 2.1 & Argentina & 1\,679 & 2.2 \\ \hline
Australia & 10\,621 & 2.1 & USA & 1\,560 & 2.0 \\ \hline
Germany & 8\,259 & 1.7 & Poland & 1\,543 & 2.0 \\ \hline
\end{tabular}
\caption{The most represented countries in the subsamples of beta testers and standard users.}
\label{tab:countries}
\end{table}

\subsection{Gender and age}

\begin{figure*}
\centering
\begin{minipage}[b]{0.5\textwidth}
\includegraphics[width=\textwidth]{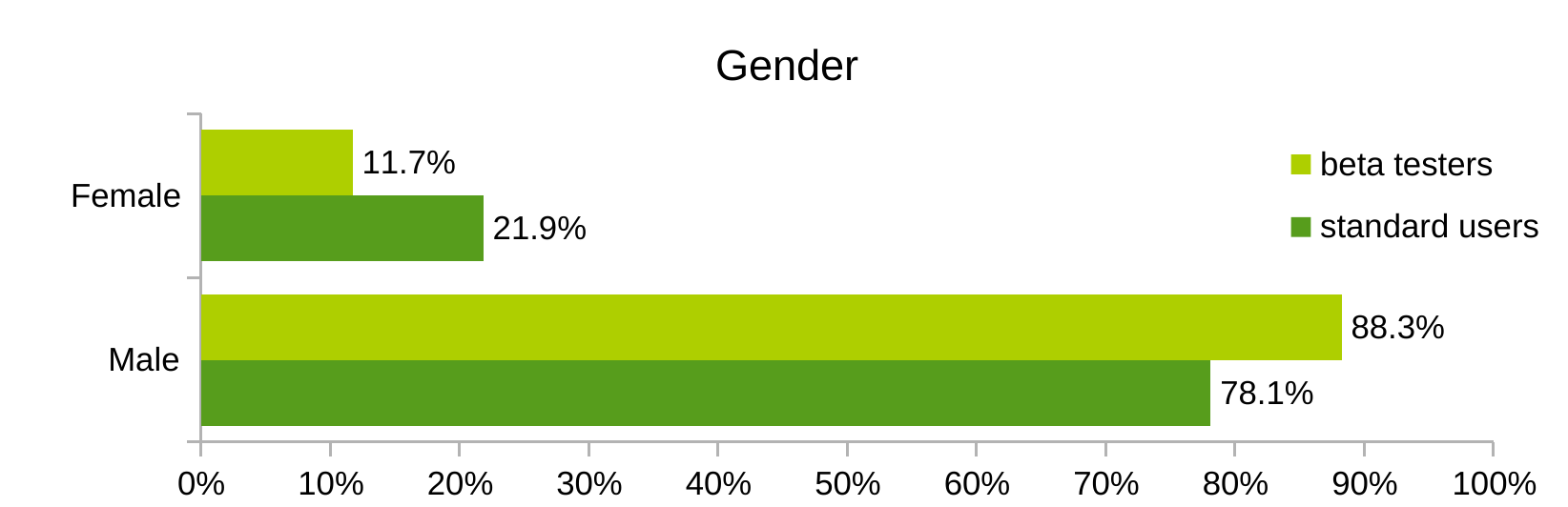}\\
\includegraphics[width=\textwidth]{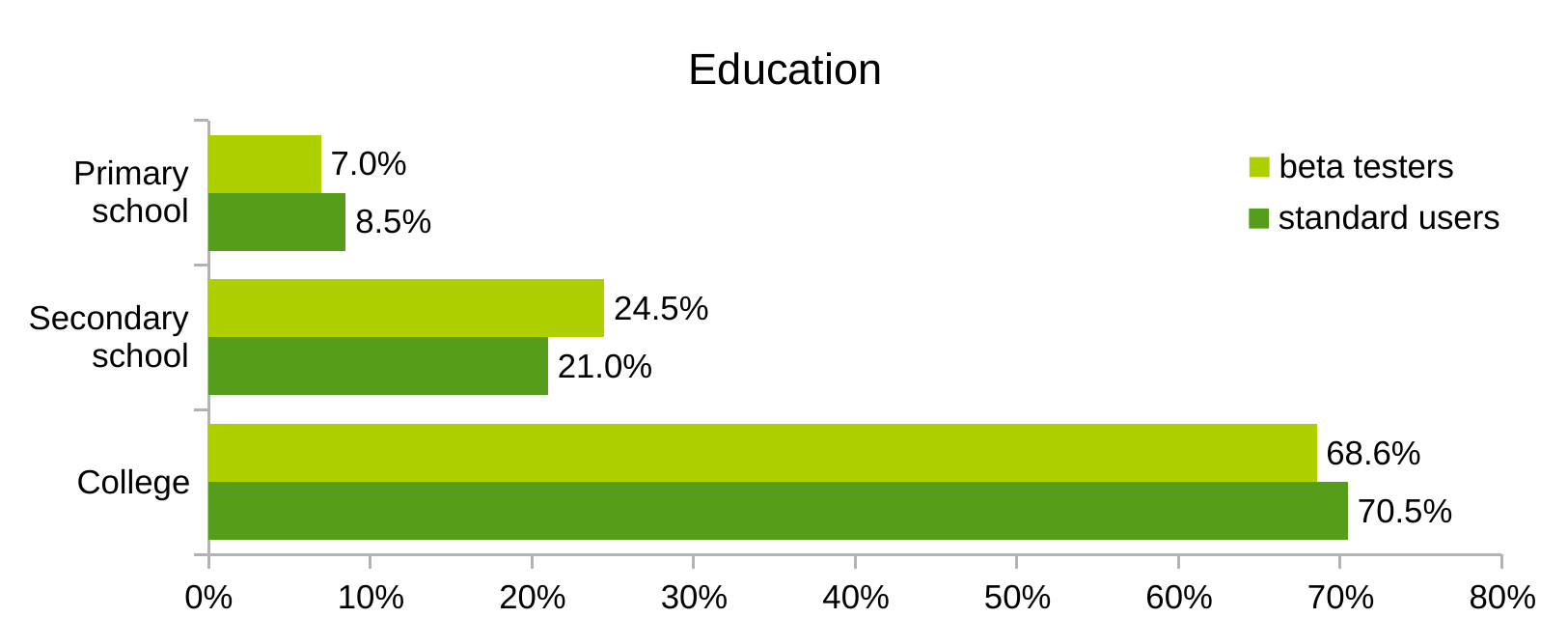}
\end{minipage}
\includegraphics[width=0.48\textwidth]{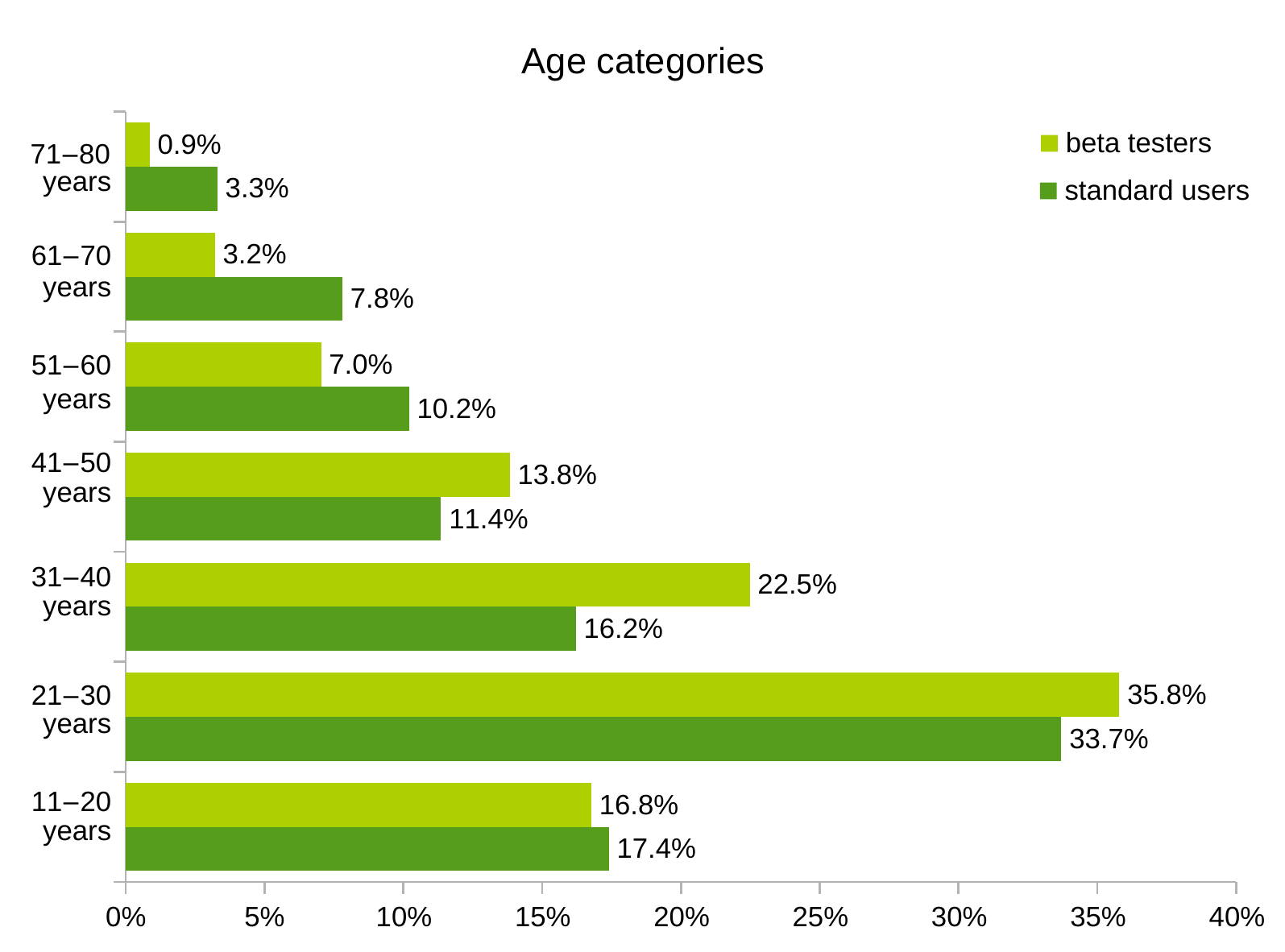}
\caption{Demographic comparison of the subsamples based on the questionnaire data.}
\label{fig:demographics}
\end{figure*}

For basic information regarding demography, see Figure~\ref{fig:demographics}. In both subsamples, males represent a vast majority, however, there are more females among standard users ($\chi^2$(1)= 277.493, $\varphi_c$ = 0.099, $p$ < 0.001, $N$ = 28\,328). 

Standard users are on average older than beta testers ($M_{beta}$ = 32.96, $SD$ = 12.974; $M_{standard}$ = 35.74, $SD$ = 16.327; $t$-test(25\,938) = 11.108; $p$ < 0.001; $d$ = 0.195, $N$ = 28\,940). Due to wide age range (11-80), we categorized the age into 7 groups to examine the differences in a more informative way (see Figure~\ref{fig:demographics}). Between the age of 21 and 50, there are significantly more beta testers than standard users, while the opposite applies to other categories ($\chi^2$(6)= 366.286, $\varphi_c$ = 0.119, $p$ < 0.001).

\subsection{Achieved education}
Education shows a consistent pattern in both subsamples, with college education being represented the most and primary school education the least. The pattern remains the same even when the youngest users (i.e., those who could not have reached higher education levels) are omitted. Beta testers are more represented in secondary education than standard users, but the effect size is small ($\chi^2$(2)= 237.085, $\varphi_c$ = 0.038, $p$ < 0.001, $N$ = 26\,354).

\subsection{Other demographic insights}
We combined the above-mentioned demographic information to examine whether beta testers represent various demographic segments of standard users well. By combining seven categories of age, gender, and education, we identified 42 unique combinations. Only two combinations were present in the sample of standard users, but not among beta testers. Both were females 71-80 years old, one with primary ($N_{standard}$ = 4) and the other with college education ($N_{standard}$ = 109). Remaining combinations were present in both subsamples with a fairly similar distribution. The highest difference was found among males aged 31-40 with college education, who were represented more often among beta testers (14.172\%) than standard users (9.539\%). 

\section{Computer self-efficacy and privacy perception}
We assessed users' computer self-efficacy and privacy perception, using dedicated questions in an optional questionnaire. Furthermore, we inspected installation-related actions such as displaying the target installation folder.

The installation included an option to change the installation folder. In order to do this users had to click on the `change installation folder' link on one of the screens during the installation process to go to respective screen. This is also the only way the user could see the default installation folder, which is not displayed otherwise. Only a minority did this, with beta testers visiting the screen more than twice as much as standard users (1.1\% of standard users and 2.6\% of beta testers). The difference is statistically significant, however, the effect size is negligible ($\chi^2$(1)= 1215.180, $\varphi_c$ = 0.046, $p$ < 0.001, $N$ = 576\,170).

\subsection{Computer self-efficacy and digital skills}

Two questions assessed users digital skills:
\begin{itemize}
\item Do you consider yourself to be a skilled computer user? (Likert scale from (1) \textit{not at all skilled} to (6) \textit{extremely skilled}.)
\item Regarding this computer, are you IT technician? (Y/N)
\end{itemize}

Beta testers are more often IT technicians ($\chi^2$(1)= 285.988, $\varphi_c$ = 0.110, $p$ < 0.001, $N$ = 23\,607) and judge themselves as more skilled than standard users ($M_{beta}$ = 4.46, $SD$ = 1.313; $M_{standard}$ = 4.18; $SD$ = 1.473; $t$-test(22\,631) = -11.743; $p$ < 0.001; $d$ = 0.200, $N$ = 22\,633).

\subsection{Privacy perception}

The last part of our questionnaire involved questions about how private data are stored in users' computers, how sensitive the users are regarding their privacy, and their beliefs about the computer being generally a safe device. All items were measured on 6-point Likert scale ranging from (1) \textit{not at all} to (6) \textit{extremely (private/sensitive/safe)}:
\begin{itemize} 
\item Do you consider the data in this computer private?
\item In general, are you sensitive about your privacy?
\item In general, do you consider computers to be safe devices against online attacks, e.g., viruses, hacking, phishing, etc.?
\end{itemize}

Both samples reported the same average level of private data in their computers ($M_{beta}$ = 4.678, $SD$ = 1.419; $M_{standard}$ = 4.690, $SD$ = 1.560; $t$-test(24\,323) = 0.504; $p$ = 0.614, $N$ = 24\,325) and both were quite similar in being privacy sensitive ($M_{beta}$ = 4.755, $SD$ = 1.376; $M_{standard}$ = 4.809; $SD$ = 1.492; $t$-test(23\,976) = 2.272; $p$ < 0.05; $d$ = 0.037, $N$ = 23\,978). We found a small difference in their evaluations of general computer safety: beta testers considered computers as slightly safer devices than standard users ($M_{beta}$ = 4.098, $SD$ = 1.712; $M_{standard}$ = 3.902; $SD$ = 1.819; $t$-test(23\,832) = -6.784; $p$ < 0.001; $d$ = 0.111, $N$ = 23\,834). We observed that beta testers consider themselves skilled IT users and they also consider the computer being a safer device than standard users do. This might suggest they are aware of security risks connected with computer usage and feel capable of preventing them. 

\section{Study limitations}

There are limitations out of our control that may have influenced the presented results. Despite the careful cleaning process, we cannot be completely sure that each record corresponds to a unique participant/device. The OS version is based on the Windows system variable \textit{CurrentVersion} that does not differentiate end user and server products. However, we presume the amount of servers in the study is negligible, as the installed ESET product is designed for end user devices. We also lacked details of devices' technological aspects, which might have shown more nuanced configuration discrepancies. 

The small ratio of users filling in the questionnaire may pose other limitations.
Firstly, self-selection and non-response bias might have skewed our results. For instance, the majority of users have college education -- these may recognize the value of user feedback better and be more willing to complete a product-related questionnaire. However, they did not differ in terms of hardware nor software from those skipping the questionnaire.
Secondly, there were only limited options to validate users' answers. Despite the thorough cleaning, some flawed questionnaire answers may have remained.
Thirdly, the questionnaire was distributed in English, which might have discouraged users not proficient in the language.

The datasets have different numbers of participants and come from different times. This may have influenced, for example, the number of people using Windows 10 as the study was conducted during the free upgrade period. Furthermore, the research was based only on the English mutation of the software, missing the customers preferring other languages.

\section{Conclusions}

We have cooperated with the software security company ESET in a large-scale comparison between beta testers and standard users of their main product. We focused on technological aspects, demographics and computer self-efficacy of nearly 600\,000 users.

Beta testers were early adopters of newer operating systems -- their distribution was significantly skewed towards newest versions (despite having less time for Windows 10 migration). They also tend to be younger, more often males, and perceive themselves as more skilled with their computers and also are more often IT technicians, supporting the `beta testers as geeks' picture. However, their hardware (platform, CPU performance and RAM size) was very similar to that of standard users, somewhat contradicting this popular image.

A striking difference was found in the countries of origin. From the top ten most represented countries only three were represented in both subsamples.

Overall, beta testers in our case study represented the population of standard users reasonably well: we have not observed any standard user segment that would be largely underrepresented in the sample of beta testers. ESET approach not to filter beta testers in any way and go with `the more testers the better', followed by analyses of selected observed differences, seems sufficient. For large international companies who are able to attract large numbers of beta testers, this may be the most efficient approach. However, for smaller, local, or not so well-established companies, this approach would probably not yield representative outcomes and may even shift the development focus in a wrong direction~\cite{dolan1993maximizing}.

\newpage
Additional resources, including the article presentation video, are available
at http://crcs.cz/papers/cacm2018.

\subsection*{Actionable takeaways}
\begin{itemize}
\item Use data you can collect to find out who your users and beta testers are. Consider the country of origin, software \& hardware configuration and basic demographics you know.
\item The fewer testers you have, the pickier you should be about their selection.
\item When testing international products, ensure beta testers are culturally representative of standard users to identify localization and cultural usability issues.
\item Testers should be representative of standard users. Keep checking that this is the case -- or remediate with additional analyses and more carefully reached conclusions.
\end{itemize}

\subsection*{Acknowledgement}
We thank Masaryk University (MUNI/M/1052/2013) and Miroslav Bartosek for support, and to anonymous reviewers and Vit Bukac for valuable feedback.

\bibliographystyle{abbrv}
\bibliography{bibliography}
\balancecolumns

\end{document}